\documentclass[conference]{IEEEtran}
\usepackage{units}

\makeatletter

\usepackage{amsfonts}
\usepackage{amsmath}
\usepackage{amssymb}
\usepackage{tikz}
\usepackage{xcolor}
\usepackage{pgfplots}
\pgfplotsset{compat=newest}

\def\as{\overset{a.s.}{\longrightarrow}}

\def\ct{\mathsf{H}}
\def\nt{\mathsf{T}}
\newcommand{\mb}[1]{
\mathbf{#1}
}
\newcommand{\TR}[1]{
\text{Tr}\left(#1\right)
}

\newtheorem{theorem}{Theorem}
\newtheorem{lemma}{Lemma}

\newtheorem{remark}{Remark}

\newtheorem{definition}{Definition}

\title{Performance Analysis of Massive MIMO Networks with Random Unitary Pilot Matrices}
\author{\IEEEauthorblockN{Rusdha Muharar}
\IEEEauthorblockA{Department of Electrical and Computer Engineering\\
Universitas Syiah Kuala, Indonesia} \and \IEEEauthorblockN{Jamie Evans} \IEEEauthorblockA{Department of
Electrical and Electronic Engineering,\\The University of Melbourne,
Australia}}

\makeatother

\begin{document}
\maketitle
\begin{abstract}
A common approach to obtain channel state information for 
massive MIMO networks is to use the same orthogonal training sequences
in each cell. We call this the full-pilot reuse (FPR) scheme. In this paper, we study an alternative approach where each cell uses different sets of orthogonal pilot (DOP) sequences. Considering uplink communications with matched filter (MF) receivers, we first derive the SINR  in the large system regime where the number of antennas at the base station, the number of users in each cell, and training duration grow large with fixed ratios. For tractability in the analysis, the orthogonal pilots are drawn from Haar distributed random unitary matrices. The resulting expression is simple and easy to compute. As shown by the numerical simulations, the asymptotic SINR approximates the finite-size systems accurately. Secondly, we derive the user capacity of the DOP scheme under a simple power control and show that it is generally better than that of the FPR scheme.
\end{abstract}

\section{Introduction}

One of the candidates for 5G technology is the massive Multiple-Input
and Multiple-Output (MIMO) system (see e.g., \cite{Boccardi_commag14,Rusek_spmag14})
introduced by Marzetta in \cite{Marzetta_wcom10}. In massive MIMO
cellular networks, a large number of small and low-cost antennas,
in the order of hundreds, is employed at base stations
(BSs). This enables an aggressive spatial multiplexing which can lead
to a ten times capacity increase compared to conventional MIMO systems
\cite{Rusek_spmag14,Marzetta_wcom10}. 
%Assuming time division duplexing (TDD) transmission with perfect channel
%reciprocity and favorable channel conditions where fast-fading channel
%coefficients are i.i.d., \cite{Marzetta_wcom10} has shown that the
%communication systems are free from the effects of fast fading and
%uncorrelated noise. This also implies that the intra-cell and inter-cell
%interferences vanish and the major cause that limits the performance
%of massive MIMO systems is the so called pilot contamination.

%This phenomenon happens because the channel estimate of a particular user contains the %channel of other cells' users that use the same (or correlated) uplink training sequence.

A common approach in the uplink training of massive MIMO systems is the
full-pilot reuse (FPR) where the same orthogonal training sequences
are used in each cell, see e.g., \cite{Marzetta_wcom10,hoydis_jsac13,Khrisnan_wcom14}.
The other approach proposed in \cite{Marzetta_wcom10} is to use different orthogonal pilots in different cells and we denote this approach as the DOP scheme. To the best of our knowledge, this scheme is largely unexplored.

It is argued in \cite{Marzetta_wcom10} that this scheme gives little difference in terms of the achieved SINRs compared to the FPR scheme. However, in the DOP, we get a small amount of contamination from all inter-cell users rather than a (potentially) large amount from a few users (those with the same pilot sequence) and this can lead to a better user capacity (see \cite{Sivamalai_globe16}). Note that the analysis in \cite{Marzetta_wcom10,Sivamalai_globe16} is performed in the regime where the number of antennas ($N$) tends to infinity and the number users ($K$) is finite ($K\ll N$). This implies that the cell-loading ($K/N$) is close to zero. It should be noted that the analysis in this regime gives a loose approximation for  finite-size systems and can also converge slowly \cite{hoydis_jsac13,Smith_icc14}.

In this paper, we generalize the performance analysis of the DOP scheme for arbitrary numbers of cell-loading. We obtain the approximation of the SINR by
performing the analysis in the large system regime where 
the number of antennas at the base station, the number of users in each
cell, and the number of training symbols go to infinity with fixed ratios. For analytical tractability, we choose the training pilots from
\textit{Haar}-distributed random unitary matrices. This approach has been used
previously in CDMA systems, see for example
\cite{Debbah_it03} and \cite{Peacock_it06}. Our numerical simulations show that the asymptotic results approximate the finite-size systems accurately.
In the analysis, we show that the pilot contamination in the asymptotic SINR expression is the \textit{average} of the square of received powers of \textit{all users} from the interfering
cells (see also \cite{Marzetta_wcom10} for a similar conclusion). This result differs from that obtained in the FPR case \cite{Marzetta_wcom10, hoydis_jsac13,Khrisnan_wcom14}, where the pilot contamination is the sum of
the received power of users from the interfering cells that use the
\textit{same} training sequence.

In this paper, we also consider another performance criterion, i.e., the user capacity. In the downlink with maximum ratio transmission (MRT) precoders, the user capacities of massive MIMO networks for single and multi-cell scenarios have been characterized in \cite{Shen_wcom15,Akbar_tcom16}, respectively. Recent work \cite{Sivamalai_globe16} studies the uplink user capacity when the cell-loading approaches zero. Here, we derive the uplink user capacity for arbitrary numbers of cell-loading under a simple power control where the uplink transmit power of a user is the inverse of the slow path-gain of that user (see also \cite{Bjornson_wcom16}). Our numerical simulations show that even though all users from the
interfering cells contribute to the pilot contamination in the DOP
scheme, its user capacity can be larger compared to that of FPR. 
Other related work is \cite{Bjornson_wcom16} that investigates the optimal number of users that maximizes the spectral efficiency of massive MIMO networks.  

The following notations are used in this paper. The boldface lower and upper case letters denote vectors and matrices, repectively. $\mathbf{I}_N$  denotes an $N\times N$ identity matrix. $\mathbb{E}[\cdot]$ and $\as$ denotes respectively the statistical expectation and the almost sure convergence. The circularly symmetric complex Gaussian (CSCG) vector with zero mean and covariance matrix $\mathbf{\Sigma}$ is denoted by $\mathcal{CN}(\mb{0},\mathbf{\Sigma})$. $|a|$ and $\Re[a]$ denote the magnitude and the real part of the complex variable $a$, respectively. $\|\cdot\|$ represents the Euclidean norm.  $\TR{\cdot}$, $(\cdot)^\nt$ and $(\cdot)^\ct$ refer to the trace, transpose and Hermitian transpose, of a matrix respectively.  
%Early works in characterizing the user
%capacity in CDMA networks, such as \cite{Tse_it99,Viswanath_it99}
%to name a few, are also relevant.

%We also performed some numerical simulations to validate the accuracy
%of the large system approximations and to show the user capacity region
%under power control for the DOP and FPR schemes.

\section{System Model}

\label{sec:system_model}

We consider a multi-cell communication system with $L$-cells. Each
cell has a base station equipped with $N$ antennas and the number
of users in cell $i$ is denoted by $K_{i}$. The channel between
user $k$ in cell $j$ and the BS in cell $i$ is denoted by the column
vector $\mathbf{g}_{kji}$ that can be modeled as 
\begin{equation}
\mathbf{g}_{kji}=\sqrt{\ell_{kji}}\mb{h}_{kji}\label{eq:ch_model}
\end{equation}
where $\mathbf{h}_{kji}\sim\mathcal{CN}(\mathbf{0},\mathbf{I}_{N})$
represents the fast-fading channel coefficients. It is assumed that
the slow-fading coefficient $\ell_{kji}$ is distance-dependent and
the shadowing effect is ignored. The channel variations follow the
block channel fading model and one block duration is equal to the channel
coherence time.

We also consider the time-division duplex (TDD) protocol with perfect channel reciprocity
between the uplink and downlink channels. In this paper, we focus
on the uplink transmission where all scheduled users transmit simultaneously
to their base station (BS). In the pilot-based TDD training, the BS estimates each 
user channel from the pilot symbols sent by each
user in the uplink transmission. The BS uses the channel information
to decode the transmitted symbols from its users. We should note that the uplink training and the uplink
data transmissions should occur in one coherence block time.

\subsection{Uplink Training\label{subsec:Uplink-Training}}
Let $T$ be the length of the channel coherence time (in symbols). Let $\tau\leq T$ be the uplink training interval or the number of training symbols. In this training phase, each user in each cell sends the pilot symbols to their BS.
%
%In this phase, for each channel block with duration $T$ (channel
%coherence time), each user in each cell sends the pilot symbols to
%their BS. Let $\tau\leq T$ be the training period or the number of
%training symbols. 
We assume a synchronized
training, where all cells perform the training at the same time and
with the same training period $\tau.$ Let $h_{kji}^{[n]}$ be the
$n$-th element of $\mathbf{h}_{kji}$ that represents the channel
from the corresponding user to the $n$-th antenna of BS $i$. Also, let $q_{ki}^{(t)}$ be the pilot symbol sent by user $k$ in cell $i$ at time $t$ and $\varrho_{ki}$ be the corresponding average training power. Note that, in the uplink training, BS $i$ will estimate the channels of its users $\mb{h}_{kii}$. Since our assumption that the elements of $\mb{h}_{kii}$ are independent, therefore we can estimate each element of the channel vector $\mb{h}_{kii}$ independently. The received signal vector at the $n$-th antenna of BS for $\tau$ training symbols is
\begin{equation}
\mathbf{y}{}_{i}^{[n]}=\sum_{j=1}^{L}\sum_{k=1}^{K_{j}}\sqrt{\rho_{kji}}h{}_{kji}^{[n]}\mathbf{q}{}_{kj}+\mathbf{n}{}_{i}^{[n]}\label{eq:rec_sig_training}
\end{equation}
where $\mathbf{y}_{i}^{[n]}=[y{}_{1i}^{[n]},\ldots,y{}_{\tau i}^{[n]}]^{\textsf{T}}$,
$\mathbf{q}_{kj}=[q_{kj}^{(1)},\ldots,q_{kj}^{(\tau)}]^{\nt}$,
and $\mb{n}_{i}^{[n]}=[n{}_{1i}^{[n]},\ldots,n_{\tau i}^{[n]}]^{\nt}\sim \mathcal{CN}\left(\mathbf{0},\sigma^{2}\mathbf{I}_{\tau}\right)$
is the receiver noise vector. We also denote $\rho_{kji}=\varrho_{kj}\ell_{kji}$
as the received power of user $k$ from cell $j$ at BS $i$.
From \eqref{eq:rec_sig_training}, we can see that BS $i$ also receives the training transmissions from other cells. It is assumed that the BS knows
all the path-gains and training sequences of all users perfectly. 
We also assume that the training matrix for cell $j$, $\mb{Q}_j=[\mb{q}_{1j},\mb{q}_{2j},\cdots,\mb{q}_{K_jj}]$ 
is obtained by extracting $K_j\leq\tau$ columns of a $\tau\times\tau$ \textit{Haar}-distributed random unitary matrix $\pmb{\mathfrak{U}}_j$ (see also \cite{Debbah_it03,Peacock_it06}). Thus, $\mb{Q}_j^\ct\mb{Q}_j=\mb{I}_K$ which implies that the training sequences are orthogonal (orthonormal) across the users in the same cell. It is assumed that $\pmb{\mathfrak{U}}_j, j=1,\ldots,L$ are independent \cite{Peacock_it06}. In other words, different cells  employs  different sets (independent) of orthogonal training sequences.  It is in contrast to the majority of works in massive MIMO 
where the same orthogonal training sequences are used in each cell. 

Let us focus on obtaining the estimate for $h_{kji}^{[n]}$. By correlating the observation
vector $\mathbf{y}_{i}^{[n]}$ with the training vector $\mathbf{q}_{ki}$, we have 
\begin{equation}
\mathbf{q}_{ki}^{\ct}\mathbf{y}_{i}^{[n]}=\sqrt{\rho_{kii}}h_{kii}^{[n]}+\sum_{j\neq i}^{L}\sum_{m=1}^{K_{j}}\sqrt{\rho_{mji}}h{}_{mji}^{[n]}\mathbf{q}_{ki}^{\ct}\mathbf{q}_{mj}+\bar{n}_{ki}^{[n]}\label{eq:obs_corr}
\end{equation}
where $\bar{n}{}_{ki}^{[n]}=\mathbf{q}_{ki}^{\ct}\mathbf{n}{}_{i}^{[n]}\sim\mathcal{CN}(0,\sigma^{2})$.
The minimum mean-square estimation (MMSE) is employed based on
the observation \eqref{eq:obs_corr}. Note that since BS $i$ knows
$\rho_{kji}$, and $\mathbf{q}_{kj}$, $\forall k,j$,
the vector $\mathbf{y}_{i}^{[n]}$ is Gaussian. Moreover, the scalars $\mathbf{q}_{ki}^\ct\mathbf{y}_{i}^{[n]}$
and $h_{kii}^{[n]}$ are jointly Gaussian. The MMSE estimate
for $h_{kii}^{[n]}$ is given by 
\begin{equation}
\hat{h}_{kii}^{[n]}=\frac{\upsilon_{ki}}{\sqrt{\rho_{kii}}}\mathbf{q}_{ki}^{\ct}\mathbf{y}_{i}^{[n]}\label{eq:ch_n_est}
\end{equation}
where 
\begin{equation}
\upsilon_{ki}=\dfrac{\rho_{kii}}{\rho_{kii}+{\displaystyle \sum_{j\neq i}^{L}\sum_{m=1}^{K_{j}}\rho_{mji}|\mathbf{q}_{ki}^{\ct}\mathbf{\mathbf{q}}_{mj}|^{2}+\sigma^{2}}}.\label{eq:var_ch_est}
\end{equation}
Note that $\frac{\upsilon_{ki}}{\sqrt{\rho_{kii}}}$  is the scalar
estimator for $h_{kii}^{[n]}$ based on the observation \eqref{eq:obs_corr}
and $\upsilon_{ki}$ is the variance of the channel estimate. It can be checked that the channel estimates for different users (in the same cell) are correlated. In contrast, they are uncorrelated in the FPR scheme. From \eqref{eq:ch_n_est}, we can model $\hat{h}_{kii}^{[n]}\sim\mathcal{CN}\left(0,\upsilon_{ki}\right)$.
Note that we have removed the index $n$ in the notation for the channel
estimation variance because it does not depend on $n$. The channel
estimate vector, $\widehat{\mathbf{h}}_{kii}=\left[\hat{h}_{kii}^{[1]},\ \hat{h}_{kii}^{[2]},\ldots,\ \hat{h}_{kii}^{[N]}\right]^{\nt}$
can be expressed as follows
\begin{align}
\mathbf{\widehat{h}}_{kii} & =\upsilon_{ki}\left(\mathbf{h}_{kii}+\sum_{j\neq i}^{L}\sum_{m=1}^{K_{j}}\sqrt{\frac{\rho_{mji}}{\rho_{kii}}}\mathbf{h}_{mji}\mathbf{q}_{mj}^{\ct}\mathbf{q}_{ki}+\frac{\hat{\mathbf{n}}_{ki}}{\sqrt{\rho_{kii}}}\right)\label{eq:ch_est_vec}
\end{align}
where $\hat{\mathbf{n}}_{ki}=[\mathbf{n}_{i}^{[1]},\cdots,\mathbf{n}_{i}^{[N]}]^{\nt}\mathbf{q}_{ki}^{*}\sim\mathcal{CN}\left(\mathbf{0},\sigma^{2}\mathbf{I}_{N}\right)$.
Note that $\mathbf{\widehat{h}}_{kii}\sim\mathcal{CN}\left(\mathbf{0},\upsilon_{ki}\mathbf{I}_{N}\right)$.
By the property of MMSE, we can model $\mathbf{h}_{kii}=\mathbf{\widehat{h}}_{kii}+\widetilde{\mathbf{h}}_{kii}$,
where $\widetilde{\mathbf{h}}_{kii}\sim\mathcal{CN}\left(\mathbf{0},(1-\upsilon_{ki})\mathbf{I}_{N}\right)$
is the channel estimation error vector.
Moreover, $\mathbf{\widehat{h}}_{kii}$ and $\widetilde{\mathbf{h}}_{kii}$
are independent. Note that the second term of \eqref{eq:ch_est_vec}
is the pilot contamination term which includes the channels of all
users from the interfering cells. In the FPR scheme, only users that
have the same training sequence contribute to this term.

\subsection{Uplink Data Transmission}

The uplink data transmission will take the duration of $T-\tau$ symbols.
Let $s_{kj}$ be the transmitted symbol from user $k$ in cell $j$ with the average power $p_{kj}$. The received signal at BS $i$
is 
\begin{equation}
\mathbf{u}_{i}=\sum_{j=1}^{L}\sum_{k=1}^{K_{j}}\sqrt{p_{kj}\ell_{kji}}\mb{h}_{kji}s_{kj}+\textsf{\textbf{n}}{}_{i}
\end{equation}
where $\textsf{\textbf{n}}{}_{i}\sim\mathcal{CN}(\mb{0},\sigma^{2}\mb{I}_{N})$ is the receiver noise at BS $i$. It is assumed that $s_{kj}\sim\mathcal{CN}(0,1)$
and is independent of other users' data symbols. For the rest of the
paper, we assume $p_{kj}=\varrho_{kj}$ and thus, $\rho_{kji}=p_{kj}\ell_{kji}$.

Let us consider user $k$ at cell $i$. A linear receiver, denoted by $\mb{c}_{ki}$, is used by BS $i$ to decode the received data symbols. The estimate for symbol $s_{ki}$ is 
\begin{align*}
\hat{s}_{ki}=\mb{c}_{ki}^{\ct}\mb{u}_{i} =\mb{c}_{ki}^{\ct}\sum_{j=1}^{L}\sum_{k=1}^{K_{j}}\sqrt{\rho_{kji}}\mb{h}_{kji}s_{kj}+\mb{c}_{ki}^{\ct}\textsf{\textbf{n}}_{i}.
%
%& =\sqrt{\rho_{kii}}\left\Vert \widehat{\mb{h}}_{kii}\right\Vert ^{2}s_{ki}+\sqrt{\rho_{kii}}\widehat{\mb{h}}_{kii}^{\ct}\widetilde{\mb{h}}_{kii}s_{ki}\\
% & \quad+\underset{(j,m)\neq(i,k)}{\sum^{L}\sum^{K_{j}}}\sqrt{\rho_{mji}}\widehat{\mb{h}}_{kii}^{\ct}\mathbf{h}_{mji}s_{mj}+\widehat{\mb{h}}_{kii}^{\ct}\textsf{\textbf{n}}{}_{i}.
\end{align*}
In this paper, we consider the MF receiver due to its low complexity. A more sophisticated LMMSE receiver is left for future works. The receiver is constructed by using the channel estimate, i.e., $\mb{c}_{ki}=\widehat{\mb{h}}_{kii}$. The resulting SINR, denoted by $\gamma_{ki}$, can be written as follows
\begin{equation}
\text{\ensuremath{\gamma}}_{ki}=\frac{{\displaystyle \rho_{kii}\|\widehat{\mathbf{h}}_{kii}\|^{4}}}{{\displaystyle \xi_{ki}+\underset{(j,m)\neq(i,k)}{\sum^{L}\sum^{K_{j}}}\rho_{mji}|\widehat{\mathbf{h}}_{kii}^{\ct}\mathbf{h}_{mji}|^{2}+\sigma^{2}\|\widehat{\mathbf{h}}_{kii}\|^{2}}}\label{eq:sinr_ul}
\end{equation}
where $\xi_{ki}=\rho_{kii}|\widehat{\mb{h}}_{kii}^{\ct}\widetilde{\mb{h}}_{kii}|^{2}$ is the self interference noise.

\section{Asymptotic Analysis}\label{sec:analysis}
In this section, we derive the asymptotic uplink SINR \eqref{eq:sinr_ul} by analyzing it in the large system regime, i.e., when $K_j, N$ and $\tau$ tend to infinity with $K_j/N$ and $K_j/\tau$ being finite constants. 
The derivation relies on results in random matrix theory \cite{Tulino_fnt04,Couillet_book11}, as well as the results on Haar-distributed random matrices (see e.g., \cite{Debbah_it03,Peacock_it06,Couillet_book11}). 

\begin{theorem}\label{thm:asinr_ul} Let $\mathbf{\Gamma}_{ji}=\text{diag}\left(\rho_{1ji},\rho_{2ji},\cdots,\rho_{K_{j}ji}\right)$. Suppose that the empirical spectral distribution (e.s.d.) of $\mathbf{\Gamma}_{ji}$ converges to a non-random distribution $F_{\mathbf{\Gamma}_{ji}}$. As $K_{j},N,\tau\to\infty$
with $\frac{K_{j}}{N}\to\alpha_{j}$ and $\frac{K_{j}}{\tau}\to\kappa_{j}$,
the uplink SINR $\gamma_{ki}$ converges almost
surely to $\bar{\gamma}_{ki}$ which is given by 
\begin{equation}\label{eq:asinr_ul}
\bar{\gamma}_{ki}=\frac{\rho_{kii}^{2}\bar{\upsilon}_{ki}}{\rho_{kii}{\displaystyle \sum_{j=1}^{L}\alpha_{j}\mathbb{E}\left[\mathsf{\Gamma}_{ji}\right]+\bar{\upsilon}_{ki}\sum_{j\neq i}^{L}\kappa_{j}\mathbb{E}\left[\mathsf{\Gamma}_{ji}^{2}\right]}}
\end{equation}
where 
\begin{equation}\label{eq:lim_chest_var}
\bar{\upsilon}_{ki}=\dfrac{\rho_{kii}}{\rho_{kii}+{\displaystyle \sum_{j\neq i}^{L}}\kappa_{j}\mathbb{E}\left[\mathsf{\Gamma}_{ji}\right]+\sigma^{2}}
\end{equation}
and $\mathsf{\Gamma}_{ji}$ is a random variable with distribution $F_{\mathbf{\Gamma}_{ji}}$.
\end{theorem}
\begin{IEEEproof}
See Appendix \ref{app:proof_asinr_ul}.	
\end{IEEEproof}
\begin{remark}
In practice, i.e. when the distribution of $F_{\mathbf{\Gamma}_{ji}}$
is difficult to obtain or $K_{j}$ is relatively small, we can replace
$\mathbb{E}\left[\mathsf{\Gamma}_{ji}\right]$ with its empirical value
 $\frac{1}{K_{j}}\sum_{k}\rho_{kji}$. 
\end{remark}

As mentioned previously, the large system analysis is performed to obtain the approximation for the uplink SINR. In practice, the parameters $K_j, N$ and $\tau$ are finite and the approximation \eqref{eq:asinr_ul} uses  the ratios of those parameters ($\alpha_j$ and $\kappa_j$). As shown in Section \ref{sec:num_sim}, \eqref{eq:asinr_ul} can approximate the finite-size systems accurately. Moreover,  we can infer from \eqref{eq:asinr_ul} that the uncertainty due to the fast fading has been removed and only that from the large-scale fading ($\rho_{kii}$) remains.

Now, let us compare \eqref{eq:asinr_ul} with the limiting SINR in the FPR scheme, denoted by $\bar{\gamma}_{ki}^{s}$,  given by  (see also \cite[eq. (23)]{hoydis_jsac13})
\begin{equation}
\bar{\gamma}_{ki}^{s}=\frac{\rho_{kii}^{2}\bar{\upsilon}_{ki}^{s}}{\rho_{kii}{\displaystyle \sum_{j=1}^{L}\alpha_{j}\mathbb{E}\left[\mathsf{\Gamma}_{ji}\right]+\bar{\upsilon}_{ki}^{s}\sum_{j\neq i}^{L}\rho_{kji}^{2}}}\label{eq:asinr_ul_sameOrth}
\end{equation}
where
\begin{equation}
\bar{\upsilon}_{ki}^{s}=\dfrac{\rho_{kii}}{\rho_{kii}+{\displaystyle \sum_{j\neq i}^{L}}\rho_{kji}+\sigma^{2}}=\dfrac{\rho_{kii}}{{\displaystyle \sum_{j=1}^{L}}\rho_{kji}+\sigma^{2}}.\label{eq:asymp_ch_est_var_sameOrth}
\end{equation}

%$\kappa_{j}\mathbb{E}\left[\mathsf{\Gamma}_{ji}\right]$ with $j\neq i$
%in \eqref{eq:asinr_ul} is replaced by the $\rho_{kji}$ (the
%interferences or received powers of users from the interfering cells that use the
%same training sequence as user $k$ at cell $i$). In this case, $\bar{\upsilon}_{ki}$
%becomes $\bar{\upsilon}_{ki}^{s}$ given by 
%
%Similarly, $\kappa_{j}\mathbb{E}\left[\mathsf{\Gamma}_{ji}^{2}\right]$
%in \eqref{eq:asinr_ul} is replaced by $\rho_{kji}^{2}$ and \eqref{eq:asinr_ul}
%reduces to $\bar{\gamma}_{ki}^{s}$ defined below

Observing the denominator of \eqref{eq:asinr_ul}, the first term
is the total interference (the intra- and inter-cell interference).
This term also appears in the denominator of \eqref{eq:asinr_ul_sameOrth}.
Thus the limiting interference power is the same for both FPR and DOP schemes. The
second term in the denominators of \eqref{eq:asinr_ul} and \eqref{eq:asinr_ul_sameOrth}
is the pilot contamination contributed by the users from the interfering cells.
However, this term takes different forms in both schemes . In DOP,
the pilot contamination is caused by all users in each interfering cell in the
form of the second moment of the interference powers. In contrast, the
pilot contamination in the FPR scheme is caused only 
by  the users with the same training pilot sequence. 

\begin{remark} In common massive MIMO setups, i.e., when $N\to\infty$
and $K_{j}$ is finite ($K_{j}\ll N$), or equivalently $\alpha_{j}\to0$,
\eqref{eq:asinr_ul} will reduce to
\begin{equation}
\lim_{\alpha\to0}\bar{\gamma}_{ki}=\rho_{kii}^{2}\left(\frac{1}{\tau}{\displaystyle \sum_{j\neq i}^{L}}{\displaystyle \sum_{m=1}^{K_{j}}\rho_{mji}^{2}}\right)^{-1}\label{eq:massMIMO_diffOrth}
\end{equation}
where we use 
$\frac{1}{K_{j}}\sum_{k}\rho_{kji}^{2}$ to represent $\mathbb{E}\left[\mathsf{\Gamma}_{ji}^{2}\right]$ in finite system dimensions. This agrees with the result in \cite{Marzetta_wcom10}. \end{remark} 

%Observing \eqref{eq:massMIMO_diffOrth}, we can see that the intra-
%and inter-cell interference terms disappear. It is in contrast to
%the large system results as shown in \eqref{eq:asinr_ul} and
%\eqref{eq:asinr_ul_sameOrth} where we still have the intra- and inter-cell
%interference terms which are scaled by the cell-loading $\alpha_{j}$.
%We can also conclude that $\gamma_{ki}^{0}>\bar{\gamma}_{ki}$. In
%another word, the performance achieved by the massive-MIMO assumption
%($\alpha\to0$) can be treated as the upper-bound of that by the large
%system analysis. However, as shown in Section \ref{sec:num_sim},
%the large system analysis provides a better SINR approximation for
%finite-size systems.

Now, let us consider a simple power control scheme where $p_{kj}=\frac{P_u}{\ell_{kjj}}$. Thus, the received power $\rho_{kjj}=P_u$ is the same for all users in cell $j$ (see also \cite{Bjornson_wcom16}). Consequently, $\mathbf{\Gamma}_{ji}$ becomes
\begin{align*}
\mathbf{\Gamma}_{ji} & =\begin{cases}
P_u\mathbf{I}_{K_{j}}, & j=i\\
P_u\bar{\mathbf{L}}_{ji}, & j\neq i
\end{cases},
\end{align*}
where $\bar{\mathbf{L}}_{ji}=\text{diag}\left(\frac{\ell_{1ji}}{\ell_{1jj}},\frac{\ell_{2ji}}{\ell_{2jj}},\cdots,\frac{\ell_{K_{j}ji}}{\ell_{K_{j}jj}}\right)$. Suppose that the e.s.d. of $\bar{\mathbf{L}}_{ji}$ converges almost surely to a deterministic distribution $F_{\bar{\mathbf{L}}_{ji}}$. Then, it is straightforward from Theorem \ref{thm:asinr_ul} to show that the limiting SINR under the power control, denoted by $\bar{\gamma}_{ki}^{p}$, is
\begin{equation}
\bar{\gamma}_{ki}^{p}=\bar{\gamma}_{i}^{p}=\frac{\bar{\upsilon}_{i}^{p}}{{\displaystyle \sum_{j=1}^{L}\alpha_{j}\mathbb{E}\left[\mathsf{\bar{L}}_{ji}\right]+\bar{\upsilon}_{i}^{p}\sum_{j\neq i}^{L}\kappa_{j}\mathbb{E}\left[\mathsf{\bar{L}}_{ji}^{2}\right]}}\label{eq:asinr_pc}
\end{equation}
where 
\begin{equation}
\bar{\upsilon}_{i}^{p}=\dfrac{1}{1+{\displaystyle \sum_{j\neq i}^{L}}\kappa_{j}\mathbb{E}\left[\mathsf{\bar{L}}_{ji}\right]+\sigma^{2}/P_u}\label{eq:asymp_ch_est_var_pc}
\end{equation}
and $\mathsf{\bar{L}}_{ji}$ is a random variable whose distribution
$F_{\bar{\mathbf{L}}_{ji}}$. 

We can see from \eqref{eq:asinr_pc} and \eqref{eq:asymp_ch_est_var_pc}
that the limiting SINR and channel estimation variance under power
control are the same for all users in cell $i$. Moreover, $\bar{\gamma}_{i}^{p}$ is deterministic since it does not depend on the random locations (or particular realizations
of the locations) of the users. Under the same power control, the
limiting SINR for the FPR scheme is 
\begin{equation}
\bar{\gamma}_{ki}^{s,p}=\frac{\bar{\upsilon}_{ki}^{s,p}}{{\displaystyle \sum_{j=1}^{L}\alpha_{j}\mathbb{E}\left[\mathsf{\bar{L}}_{ji}\right]+\bar{\upsilon}_{ki}^{s,p}\sum_{j\neq i}^{L}\frac{\ell_{kji}^{2}}{\ell_{kjj}^{2}}}}\label{eq:asinr_fpr_pc}
\end{equation}
where $\bar{\upsilon}_{ki}^{s,p}=\left(1+\sum_{j\neq1}^{L}\nicefrac{\ell_{kji}}{\ell_{kjj}}+\sigma^{2}/P_u\right)^{-1}$.
It is obvious that $\bar{\gamma}_{ki}^{s,p}$ still depends on the
particular realizations of $\bar{\ell}_{kji}=\nicefrac{\ell_{kji}}{\ell_{kjj}}$. Note
that, for a particular realization of $\bar{\ell}_{kji}$,
$\sum_{j\neq1}^{L}\bar{\ell}_{kji}$ and $\sum_{j\neq i}^{L}\bar{\ell}_{kji}^2$
can be larger or smaller than $\sum_{j\neq i}^{L}\kappa_{j}\mathbb{E}\left[\mathsf{\bar{L}}_{ji}\right]$
and $\sum_{j\neq i}^{L}\kappa_{j}\mathbb{E}\left[\mathsf{\bar{L}}_{ji}^{2}\right]$,
respectively. Consequently, $\bar{\gamma}_{ki}^{s,p}$ can be smaller
or larger than $\bar{\gamma}_{i}^{p}$. 

\section{User Capacity Under Power Control\label{sec:User-Capacity}}

Here we will characterize the user capacity based on the limiting
SINR under power control \eqref{eq:asinr_pc}.
Recall that the limiting SINR, $\bar{\gamma}_{i}^{p}$, is deterministic
and is the same for all users in cell $i$. Now, let us assume that
all users in cell $i$ have the quality of service (QoS) requirement, 

\begin{equation}
\bar{\gamma}_{i}^{p}\geq\gamma_{i}^{\mathsf{th}}\label{eq:QoS_up}
\end{equation}
where $\gamma_{i}^{\mathsf{th}}$ is the minimum required SINR for users in
cell $i$. Then, the user capacity is defined as  the number of users per
degree of freedom ($N$), or $\alpha_{i}$, that satisfies the 
QoS \eqref{eq:QoS_up}, see also \cite{Shen_wcom15,Akbar_tcom16}. From \eqref{eq:QoS_up}, we have
\begin{equation}
\alpha_{i}+{\displaystyle \sum_{j\neq i}^{L}}\alpha_{j}\mathbb{E}\left[\mathsf{\bar{L}}_{ji}\right]\leq\bar{\upsilon}_{i}^{p}\left(\frac{1}{\gamma_{i}^{\mathsf{th}}}-\sum_{j\neq i}^{L}\kappa_{j}\mathbb{E}\left[\mathsf{\bar{L}}_{ji}^{2}\right]\right).\label{eq:user_cap}
\end{equation}

To get more insights on the user capacity, let us consider the simplest case
where $K_{j}=K$, so that $\alpha_{j}=\alpha$, $\kappa_{j}=\kappa$. Note that $K$ affects both  $\alpha$ and $\kappa$ and we can write $\alpha=\frac{K}{\tau}\frac{\tau}{N}=\kappa\theta$ with $\theta=\frac{\tau}{N}$.
For a given degree of freedom, there are two ways of defining $\alpha$ that satisfies the QoS in \eqref{eq:QoS_up}, either by fixing $\tau$ or $\kappa$. 

\subsection{Fixed $\tau$}
Recall that $\alpha=\kappa\theta$. By substituting the expression for $\bar{\upsilon}_{i}^{p}$
in \eqref{eq:asymp_ch_est_var_pc} into \eqref{eq:user_cap}, we obtain
\begin{equation}
A_{i}\kappa^{2}+B_{i}\kappa-\frac{1}{\gamma_{i}^{\mathsf{th}}}\leq0\label{eq:quadratic_kappa_sameK}
\end{equation}
where 
\begin{align*}
A_{i} & =\theta\left({\displaystyle \sum_{j=1}^{L}}\mathbb{E}\left[\mathsf{\bar{L}}_{ji}\right]\right)\left({\displaystyle \sum_{j\neq i}^{L}}\mathbb{E}\left[\mathsf{\bar{L}}_{ji}\right]\right)\\
B_{i} & =\theta\left(1+\sigma^{2}/P_u\right){\displaystyle \sum_{j=1}^{L}\mathbb{E}\left[\mathsf{\bar{L}}_{ji}\right]+\sum_{j\neq i}^{L}\mathbb{E}\left[\mathsf{\bar{L}}_{ji}^{2}\right]}.
\end{align*}
Now we need to find the range of $\kappa$ (or $\alpha$) such that
\eqref{eq:quadratic_kappa_sameK} holds with $A_{i},B_{i},\gamma_{i}^{\mathsf{th}}>0$
and $0\leq\kappa\leq1$. The left hand side of \eqref{eq:quadratic_kappa_sameK}
is a quadratic equation with one positive root and one negative root.
The positive root (solution) is 
\begin{equation}\label{eq:kappa_tau}
\kappa_{\tau}=\frac{-B_{i}+\sqrt{B_{i}^{2}+\frac{4A_{i}}{\gamma_{i}^{\mathsf{th}}}}}{2A_{i}}.
\end{equation}
It is easy to check that the left hand side of \eqref{eq:quadratic_kappa_sameK} is a convex function and
has a negative value at $\kappa=0.$ Thus, \eqref{eq:quadratic_kappa_sameK}
will hold when 
\[
0\leq\kappa\leq\kappa_{\tau}.
\]
Equivalently, the cell loading that satisfies \eqref{eq:QoS_up} is
\begin{equation}
0\leq\alpha\leq\alpha_{\tau},\quad\text{with }\alpha_{\tau}=\kappa_{\tau}\theta.\label{eq:alpha_tau}
\end{equation}
Now, we will check the condition where $\kappa_{\tau}\leq1$ holds.
It is equivalent to $\sqrt{B_{i}^{2}+\frac{4A_{i}}{\gamma_{i}^{\mathsf{th}}}}\leq2A_{i}+B_{i}$.
By squaring both sides and performing some algebraic manipulations, $\kappa_{\tau}\leq1$
holds when
\[
\frac{1}{\gamma_{i}^{\mathsf{th}}}\leq A_{i}+B_{i}.
\]
Otherwise, we set $\kappa_{\tau}=1$ (or $\alpha_{\tau}=\theta$). 

Considering  \eqref{eq:kappa_tau} and \eqref{eq:alpha_tau}, it can be easily checked that $\alpha_\tau$ is an increasing function of $\tau$.
Thus, a larger $\tau$ will increase the user capacity.

\subsection{Fixed $\kappa$}

In this case, $\kappa$ is fixed  and thus $\tau$ varies
as $K$ varies. The corresponding user capacity is simply given by

\begin{equation}
\alpha\leq\alpha_{\kappa}=\bar{\upsilon}_{i}^{p}\dfrac{\frac{1}{\gamma_{i}^{\mathsf{th}}}-\kappa\sum_{j\neq i}^{L}\mathbb{E}\left[\mathsf{\bar{L}}_{ji}^{2}\right]}{1+\sum_{j\neq i}^{L}\mathbb{E}\left[\mathsf{\bar{L}}_{ji}\right]}.\label{eq:alpha_kappa}
\end{equation}
Since $\alpha_{\kappa}\geq0$, the right-hand side of the above equality
must satisfy 
\[
\frac{1}{\gamma_{i}^{\mathsf{th}}}\geq\kappa\sum_{j\neq i}^{L}\mathbb{E}\left[\mathsf{\bar{L}}_{ji}^{2}\right].
\]
It is obvious that the choice of $\kappa$ affects the user capacity.
It can be checked that $\bar{\upsilon}_{i}^{p}\left(\frac{1}{\gamma_{i}^{\mathsf{th}}}-\kappa\sum_{j\neq i}^{L}\mathbb{E}\left[\mathsf{\bar{L}}_{ji}^{2}\right]\right)$
is increasing as $\kappa$ decreases. Note that by decreasing $\kappa$,
we have a better quality of channel estimates that obviously increases
the user capacity.

\begin{remark}
As discussed in both subsections above, increasing $\tau$  (or decreasing $\kappa$) will result in a higher user capacity. On the other side, this will reduce the uplink transmission time and consequently the cell sum rate. This trade-off is also a concern raised in \cite{Marzetta_wcom10}. One way to handle this trade-off is to find an optimal $\tau$, for $K\leq\tau\leq T$, that maximizes the cell sum rate. It is obvious that $\tau=T$ is never an optimal solution since it gives a zero sum-rate. Thus, the optimal $\tau$ is either at $\tau=K$ or at $\tau\in (K,T)$. Let us consider the worst case in terms of the user capacity, i.e., $\tau=K$ (or $\kappa=1$). The user capacity in this case is easily obtained from  \eqref{eq:alpha_kappa} (or \eqref{eq:alpha_tau}) and is given by 
	\[
		\alpha_{\kappa}=\dfrac{\frac{1}{\gamma_{i}^{\mathsf{th}}}-\sum_{j\neq i}^{L}\mathbb{E}\left[\mathsf{\bar{L}}_{ji}^{2}\right]}{\left(1+\sum_{j\neq i}^{L}\mathbb{E}\left[\mathsf{\bar{L}}_{ji}\right]\right)\left(1+{\sum_{j\neq i}^{L}}\mathbb{E}\left[\mathsf{\bar{L}}_{ji}\right]+\frac{\sigma^{2}}{P_u}\right)}.
	\]
	The numerical simulation for this case is shown in Figure \ref{fig:user_cap_compar2} in the following section. We can see that the user capacity of the DOP scheme is considerably larger compared to that of the FPR scheme.  
\end{remark}

\section{Numerical Simulations}

\label{sec:num_sim}

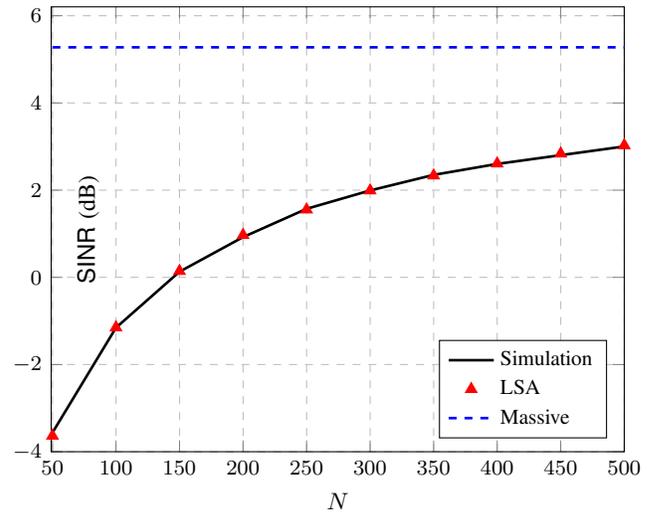
\begin{figure}
\centering 	
\begin{tikzpicture} 	
	\pgfplotsset{grid style=dashed} 	
	\begin{axis}[small, width=9.2cm, height=7.5cm, y label style={at={(axis description cs:0.1,.5)},anchor=south}, xlabel=$N$, ylabel=$\textsf{SINR}\ \text{(dB)} $,  	xmin=49, xmax=500, ymin=-4, grid=major, legend pos=south east, legend cell align=left] 	
		\addplot[no marks, line width=1]table[x=N,y=sim]{sinr_vs_N_boundedPL_exp2.dat};
		\addplot[color=red,only marks, mark=triangle*, mark size=2.5]table[x=N,y=asymp]{sinr_vs_N_boundedPL_exp2.dat};
		\addplot[color=blue, dashed, no marks, line width=1] table[x=N, y=massive] {sinr_vs_N_boundedPL_exp2.dat};
		%\addplot[color=brown, only marks, mark=*, mark size=2] table[x=N, y=asymp_pl] {Matlab Code/%sinr_vs_N_boundedPL_exp2.dat};
		\legend{\footnotesize{Simulation}, \footnotesize{LSA}, \footnotesize{Massive}} 	
	\end{axis} 
\end{tikzpicture} 
\caption{The asymptotic SINR obtained from the simulation, large system analysis (LSA) and the conventional approach ($N\to\infty$ but $K$ is finite) in massive MIMO systems.} 	
\label{fig:verify_sinr_mf}
\end{figure}

In this section, we will validate the approximation of our large system
results for finite $K_{j}$, $N$, and $\tau$. We consider a 7\textendash cell
hexagonal layout where the inner cell radius is normalized to $0.9$ and the distance between the base stations
is normalized to $2$. The users are uniformly
distributed across the cell.  We adopt the bounded path-loss model where the slow path-gain is modeled as $\ell_{kji}=(1+d_{kji}^{\zeta})^{-1}$
where $\zeta$ is the path-loss exponent and $d_{kji}$ is the
distance between user $k$ in cell $j$ and BS $i$. We set $\zeta=3.7$ and 
 ignore the shadowing effect in the model. The experiment takes
parameters $P_u=0$ dB, $K_{j}=K=20$, $\kappa=\frac{2}{3}$ and the number of
antennas varies from $50$ to $500$ with interval $50$. We generated
$500$ channel realizations according to the Rayleigh distribution.
Figure \ref{fig:verify_sinr_mf} shows the SINR (in decibels) as the
number of antennas at the BS increases. We can see that the SINR obtained
from the large system analysis or LSA \eqref{eq:asinr_ul} can approximate
the simulation results (finite-size systems, i.e., with $K=20$) accurately.
Observe that the SINR obtained by using  the conventional approach in massive MIMO ($K<\infty$ but $N\to\infty$), acts as the upper-bound for the finite-size systems. The figure also
implies that this approach gives  quite a  loose approximation compared
to the large system approach (see also \cite{Smith_icc14} for the
convergence behavior of massive MIMO systems). Observe that for $N=100$,
the SINR gap between the massive MIMO approximation
and the finite size result is about 9 dB. This gap reduces as $N$
increases and it is approximately $2.3$ dB when $N=500$. 

\begin{figure}
\centering 	
\begin{tikzpicture} 	
\pgfplotsset{grid style=dashed} 	
\begin{axis}[small, width=9.2cm, height=7.5cm, y label style={at={(axis description cs:0.1,.5)},anchor=south}, xmin=-8,xmax=4.01, ymin=0,ymax=1,grid=major, xlabel=SINR (db), ylabel=CDF] 	
\addplot[color=blue, no marks, line width=0.75]table[x=sinr,y=cdf]{cdf_same_alpha0p1_exp1.dat};
\addplot[color=red,no marks, dashed, line width=1]table[x=sinr,y=cdf]{cdf_dif_alpha0p1_kappa1p3_exp1.dat};
\addplot[color=black, dashed, no marks, line width=1] table[x=sinr, y=cdf] {cdf_dif_alpha0p1_kappa2p3_exp1.dat};
\addplot[color=brown, dashed, no marks, line width=1] table[x=sinr, y=cdf] {cdf_dif_alpha0p1_kappa1_exp1.dat};
\draw[->,thick,>=stealth] (axis cs:-2,0.7) -- (axis cs:1.14,0.7); 		\node at (axis cs:-3,0.7) {\footnotesize{$\kappa=\frac{2}{3}$}}; 	
\draw[->,thick,>=stealth] (axis cs:-2,0.6) -- (axis cs:0.14,0.6); 		\node at (axis cs:-3,0.6) {\footnotesize{$\kappa=1$}};
\draw[->,thick,>=stealth] (axis cs:-2,0.8) -- (axis cs:2.4,0.8); 		\node at (axis cs:-3,0.8) {\footnotesize{$\kappa=\frac{1}{3}$}};
%\legend{\footnotesize{$\bar{\gamma}_i^{s,p}$}, \footnotesize{$\bar{\gamma}_i^{s,p}$}}; 	
\end{axis} 
\end{tikzpicture} 

\caption{CDF of $\bar{\gamma}_{ki}^{s,p}$ (solid) and $\bar{\gamma}_{i}^{p}$
(dashed) for $\alpha=0.1$ and $\kappa=\{\frac{1}{3},\frac{2}{3},1\}$ }
\label{fig:cdf_comparison}
\end{figure}
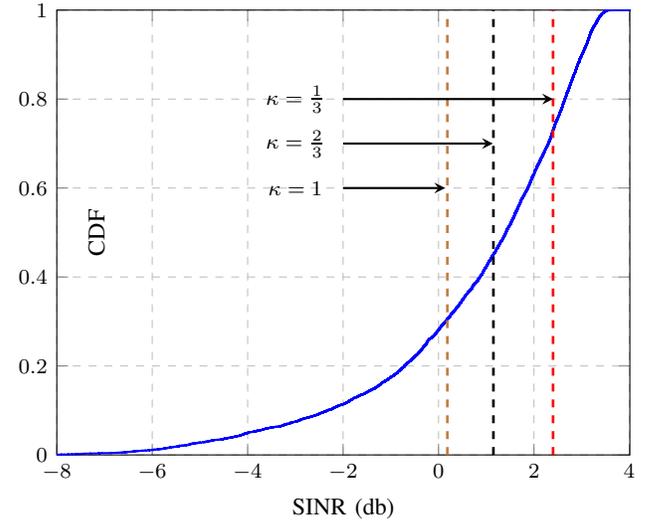

Figure \ref{fig:cdf_comparison} compares the cdf of $\bar{\gamma}_{ki}^{s,p}$
and $\bar{\gamma}_{i}^{p}$. Note that $\bar{\gamma}_{i}^{p}$ is
a deterministic quantity. Hence, its cdf is a unit impulse located
at $\bar{\gamma}_{i}^{p}$. To produce the curves, we set $\alpha=0.1$
and $\kappa=\{\frac{1}{3},\frac{2}{3},1\}$. Note that, $\bar{\gamma}_{i}^{p}$
depends on $\kappa$ while $\bar{\gamma}_{ki}^{s,p}$ does not. It
is obvious from the curves for $\bar{\gamma}_{i}^{p}$ that smaller
$\kappa$ gives a better performance (SINR). The figure also reveals
that there is some probability that $\bar{\gamma}_{i}^{p}$ is larger
than $\bar{\gamma}_{ki}^{s,p}$ and vice-versa. For example, when
$\kappa=1$, there is a $30\%$ chance that $\bar{\gamma}_{i}^{p}$
is larger than $\bar{\gamma}_{ki}^{s,p}$. When smaller $\kappa=\frac{1}{3}$
is used, the probability becomes higher i.e., around $75\%$. This
probability also changes when $\alpha$ varies. Thus, it is interesting
to see how the user capacity (which is related to $\alpha$) is affected
by both schemes. 

The user capacity for the different orthogonal training sequences
(DOP scheme) has been presented in Section \ref{sec:User-Capacity}.
For the FPR scheme, the formulation of user capacity is slightly different
since $\bar{\gamma}_{ki}^{s,p}$ is a random quantity contributed
by the large-scale fading terms of the pilot contamination. Hence,
the user capacity region, for the simplest case ($K_{j}=K,\forall j$)
is defined as the value of $\alpha$ that satisfies

\begin{equation}
\mathbb{P}\left(\bar{\gamma}_{ki}^{s,p}\geq \gamma_{i}^{\mathsf{th}}\right)\geq1-\beta\label{eq:user_cap_same}
\end{equation}
 where $\beta$ is the outage probability. The above is equivalent
to $\mathbb{P}\left(X_{ki}\leq0\right)\geq1-\beta$ or $F_{X_{ki}}(0)\geq1-\beta$
where 
\[
X_{ki}=\alpha{\displaystyle \sum_{j=1}^{L}\mathbb{E}\left[\mathsf{\bar{L}}_{ji}\right]\left(1+\frac{\sigma^{2}}{P_u}+\sum_{j\neq i}^{L}\bar{\ell}_{kji}\right)+\sum_{j\neq i}^{L}\bar{\ell}_{kji}^{2}}-\frac{1}{\gamma_{i}^{\mathsf{th}}}.
\]

\begin{figure}
\centering 	
\begin{tikzpicture} 	
\pgfplotsset{compat=newest,grid style=dashed} 	
\begin{axis}[small, width=9 cm, height=7.5cm,  xmin=-10,xmax=10, ymin=0,ymax=2,grid=major, xlabel={$\gamma_{i}^{\mathsf{th}}$ (dB)}, ylabel=User Capacity (per Antenna)  - $\alpha_{\tau}$, legend style={font=\footnotesize},legend cell align=left] 
	
\addplot[color=blue, no marks, line width=1, forget plot]table[x=sinr,y=alpha]{userCap_qos0p95.dat};

%\addplot[color=purple,no marks, line width=1, forget plot]table[x=sinr,y=alpha]{Matlab Code MF/userCap_qos0p9.dat};
%\addplot[color=black, no marks, line width=1] table[x=sinr, y=alpha] {Matlab Code/%userCap_qos0p5.dat};

\addplot[color=red, dashed, every mark/.append style={solid}, mark=*, mark size=1.5, line width=0.5] table[x=sinr, y=alpha] {userCap_dif_theta1p5.dat};

\addplot[color=brown, dashed, every mark/.append style={solid}, mark=diamond*, mark size=2, line width=0.5] table[x=sinr, y=alpha] {userCap_dif_theta1.dat};

\addplot[color=black,dashed,every mark/.append style={solid}, mark=triangle*, mark size=1.5, line width=0.5] table[x=sinr, y=alpha] {userCap_dif_theta0p5.dat};

%\draw[->,thick,>=stealth] (axis cs:-4,1.3) -- (axis cs:-7.6,0.7); 		\node[right] at (axis cs:-4,1.35) {\footnotesize{$\beta=0.1$}}; 	
%\draw[->,thick,>=stealth] (axis cs:-4,0.95) -- (axis cs:-6.9,0.42); 		\node[right] at (axis cs:-4,1) {\footnotesize{$\beta=0.05$}}; 
%\draw[->,thick,>=stealth] (axis cs:-2,0.8) -- (axis cs:2.4,0.8); 		\node at (axis %cs:-3,0.8) {\footnotesize{$\kappa=\frac{1}{3}$}};
\legend{$\theta=1.5$,$\theta=1$,$\theta=0.5$}	
\end{axis} 
\end{tikzpicture} 

\caption{User Capacity of the FPR  (solid) and the DOP (mark/dashed) schemes  for different values of $\theta=\frac{\tau}{N}$.}
\label{fig:user_cap_compar1}
\end{figure}
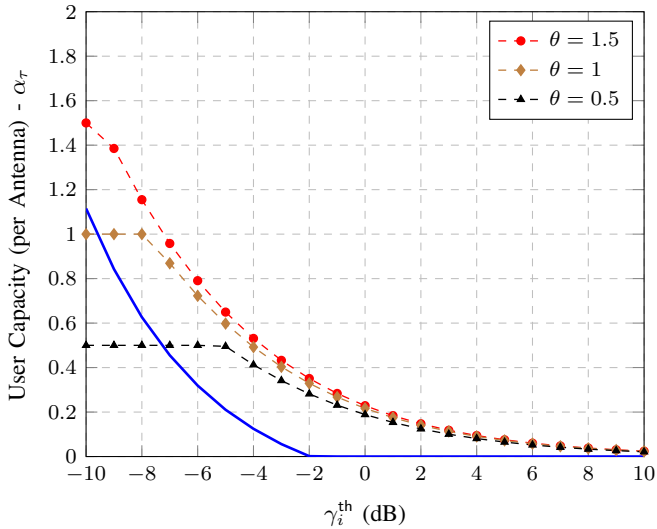

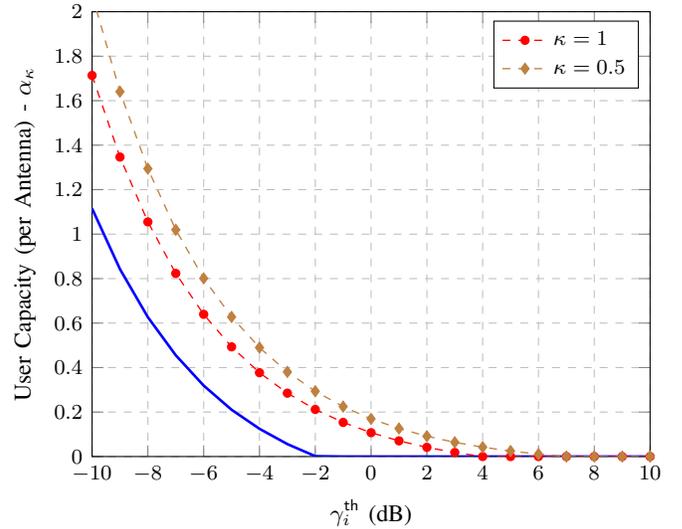
\begin{figure}
\centering 	
\begin{tikzpicture} 	
\pgfplotsset{compat=newest,grid style=dashed} 	
\begin{axis}[small, width=9 cm, height=7.5cm,  xmin=-10,xmax=10, ymin=0,ymax=2,grid=major, xlabel={$\gamma_{i}^{\mathsf{th}}$ (dB)}, ylabel=User Capacity (per Antenna) - $\alpha_{\kappa}$, legend style={font=\footnotesize},legend cell align=left] 
	
\addplot[color=blue, no marks, line width=1, forget plot]table[x=sinr,y=alpha]{userCap_qos0p95.dat};

%\addplot[color=purple,no marks, line width=1, forget plot]table[x=sinr,y=alpha]{Matlab Code MF/userCap_qos0p9.dat};
%\addplot[color=black, no marks, line width=1] table[x=sinr, y=alpha] {Matlab Code/%userCap_qos0p5.dat};

\addplot[color=red, dashed, every mark/.append style={solid}, mark=*, mark size=1.5, line width=0.5] table[x=sinr, y=alpha] {userCap_dif_kappa1.dat};

\addplot[color=brown, dashed, every mark/.append style={solid}, mark=diamond*, mark size=2, line width=0.5] table[x=sinr, y=alpha] {userCap_dif_kappa0p5.dat};

%\addplot[color=black,densely dotted,every mark/.append style={solid}, mark=triangle*, mark size=1.5, line width=0.5] table[x=sinr, y=alpha] {Matlab Code MF/userCap_dif_kappa0p2.dat};

%\draw[->,thick,>=stealth] (axis cs:-4,1.4) -- (axis cs:-8,0.8); 		\node[right] at (axis cs:-4,1.45) {\footnotesize{$\beta=0.1$}}; 	
%\draw[->,thick,>=stealth] (axis cs:-4,1.02) -- (axis cs:-7.4,0.52); 		\node[right] at (axis cs:-4,1.05) {\footnotesize{$\beta=0.05$}}; 
%\draw[->,thick,>=stealth] (axis cs:-2,0.8) -- (axis cs:2.4,0.8); 		\node at (axis %cs:-3,0.8) {\footnotesize{$\kappa=\frac{1}{3}$}};
\legend{$\kappa=1$,$\kappa=0.5$}	
\end{axis} 
\end{tikzpicture} 

\caption{User Capacity of the FPR (solid) and the DOP (mark/dashed) schemes for different values of $\kappa$.}
\label{fig:user_cap_compar2}
\end{figure}

Figure \ref{fig:user_cap_compar1} presents the comparisons of the
user capacities of the FPR obtained from \eqref{eq:user_cap_same} with $\beta=0.05$
and of the DOP represented by $\alpha_{\tau}$ in \eqref{eq:alpha_tau} with $\theta=\{0.5,1,1.5\}$.
As expected, higher outages lead to higher user capacities for the
FPR while higher $\theta$ (higher $\tau$ with $N$ fixed) gives
higher user capacities for DOP. In general, we can see that the FPR
has a higher user capacity compared to the DOP for very low $\gamma_{i}^{\mathsf{th}}$ (less than $-6$ dB)
and the reverse occurs for low to high $\gamma_{i}^{\mathsf{th}}$s. The user capacity of the FPR scheme is zero starting at  $\gamma_{i}^{\mathsf{th}}=-2\text{ dB}$ for outage probability $5\%$ and at $\gamma_{i}^{\mathsf{th}}=0\text{ dB}$.
On the other hand, the user capacities for the DOP scheme are about $0.28-0.35$ and $0.2$ when  $\gamma_{i}^{\mathsf{th}}=-2\text{ dB}$ and $\gamma_{i}^{\mathsf{th}}=0\text{ dB}$, respectively. This shows the advantage of the DOP scheme over the FPR scheme in terms of user capacity and is consistent with the findings in \cite{Sivamalai_globe16}.
A similar conclusion can be drawn by comparing the user capacities of
FPR and $\alpha_{\kappa}$ of DOP with $\kappa=\{0.5,1\}$ as shown in 
Figure \ref{fig:user_cap_compar2}. Even for the worst case, $\kappa=1$,
the user capacities of the DOP are higher than those of the FPR at all considered values of $\gamma_{i}^{\mathsf{th}}$. For example, at $\gamma_{i}^{\mathsf{th}}=-10$ dB, the user capacities in the FPR scheme and the DOP scheme are about $1.1$ and $1.7$, respectively. Hence, for example, if the BS has $500$ antennas, the DOP scheme can admit $300$ more users than the FPR scheme can.

\section{Conclusion}

\label{sec:conclusion} In this paper, we consider the performance analysis of uplink massive MIMO systems with different sets of orthogonal training pilots in different cells employed during the uplink training. We obtained a novel expression for the asymptotic SINR in the large system regime that approximates finite-size systems accurately. We show that under a simple power control, the asymptotic SINR is a deterministic quantity and its denominator contains the average interference and mean-squared interference (pilot contamination) of all users from the interfering cells. We also derived the user capacity of the system and compared to that of the FPR scheme. Our simulation results showed that   the DOP generally gives a better user capacity even for  low values of the required SINRs.
\appendices

\section{Proof of Theorem \ref{thm:asinr_ul}}

\label{app:proof_asinr_ul} 

We begin the proof by establishing the limiting result for the channel estimate variance $\upsilon_{ki}$. Then, we proceed to obtain the limiting results for the numerator and the denominator of the uplink SINR $\gamma_{ki}$ which are denoted by $A_{ki}$ and $B_{ki}$ respectively. In the proof we use the following definitions and results from large random matrices.
\begin{definition}[{\cite[Definition 2.6]{Peacock_thesis05}}] 
	Two infinite sequences $a_{n}$ and $b_{n}$ are defined to be asymptotically equivalent as $n\to\infty$, denoted by $a_{n}\asymp b_{n}$, if only if $|a_{n}-b_{n}|\as 0$.
\end{definition}

\begin{definition}[{\cite[Definition 2.7]{Peacock_thesis05}}] Two infinite sequences $a_{k,n}$ and $b_{k,n}$, indexed by $k=1,2,\cdots,n$, are defined to be uniformly asymptotically equivalent as $n\to\infty$, denoted by $a_{k,n}\overset{k}{\asymp}b_{k,n}$, if only if $\underset{k\leq N}{\max}|a_{k,n}-b_{k,n}|\as0$.
\end{definition}

\begin{lemma}[{\cite[Lemma 2.5]{Peacock_thesis05}}]\label{lemma:av_seq_lim} If $a_{k,n}\overset{k}{\asymp}b_{k,n}$, then $\frac{1}{n}\sum_{k=1}^{n}a_{k,n}\asymp\frac{1}{n}\sum_{k=1}^{n}b_{k,n}$.	
\end{lemma}

\begin{lemma}[{\cite[Lemma 1]{Evans_it00}}]\label{lemma:Lemma1_jse} Let $\mathbf{A}\in\mathbb{C}^{N\times N}$ be a deterministic matrix with uniformly bounded spectral radius for all $N$. Let $\mathbf{x}=\frac{1}{\sqrt{N}}[x_{1},x_{2},\cdots,x_{N}]^{\nt}$ where the $x_{i}$'s are i.i.d with zero mean, unit variance and finite eighth moment. Let $\mathbf{y}$ be a similar vector independent of $\mathbf{x}$. Then, we have $\mathbf{xAx}^{\ct}-\frac{1}{N}\text{Tr}(\mathbf{A})\as0$, and $\mathbf{xAy}^{\ct}\as0$.\end{lemma}

The following lemma which involves the columns of an isometric matrix is analogues to Lemma \ref{lemma:Lemma1_jse}.
\begin{lemma}\label{lemma:quadratic_iso}
	Let $\mathbf{w}\in \mathbb{C}^{N}$ be a column of an isometric matrix\footnote{An $N\times K$ isometric matrix with $K\leq N$ is obtained by taking $K$ columns of an $N\times N$ \textit{Haar}-distributed random matrix \cite[Definition 2.14]{Peacock_thesis05}.} $\mathbf{W}$ and $\mathbf{z}$ be a column of an isometric $\mathbf{Z}$ matrix independent of $\mathbf{W}$. Let $\mathbf{A}$ be as defined in Lemma \ref{lemma:Lemma1_jse} and independent of $\mathbf{W}$ and $\mathbf{Z}$. Then, $\mathbf{w}^\ct\mathbf{A}\mathbf{w}-\frac{1}{N}\TR{\mathbf{A}}\as 0$  and  $\mathbf{w}^\ct\mathbf{A}\mathbf{z}\as 0$.  
\end{lemma}

In the following subsections, we use the notation $\mathbf{X}_{[k]}$ for matrix $\mathbf{X}$ with $k$-th column removed. If $\mathbf{X}$ is a diagonal matrix, $\mathbf{X}_{[k]}$ denotes  $\mathbf{X}$ with both $k$-th column and $k$-th row deleted.

\subsection{The large system limit for $\upsilon_{ki}$}

Here, we only need to find the asymptotic limit for the second term
in the denominator of $\upsilon_{ki}$ in \eqref{eq:var_ch_est}.
That term can be written as $\sum_{j\neq i}^{L}\mathbf{q}_{ki}^{\ct}\mathbf{Q}_{j}\mathbf{\Gamma}_{ji}\mathbf{Q}_{j}^{\ct}\mathbf{q}_{ki}$.
Since $\mathbf{q}_{ki}$ is independent of $\mathbf{Q}_{j}$ for $j\neq i$,
$\mathbf{q}_{ki}^{\ct}\mathbf{Q}_{j}\mathbf{\Gamma}_{ji}\mathbf{Q}_{j}^{\ct}\mathbf{q}_{ki}\asymp\frac{1}{\tau}\text{Tr}\left(\mathbf{Q}_{j}\mathbf{\Gamma}_{ji}\mathbf{Q}_{j}^{\ct}\right)$ by Lemma \ref{lemma:quadratic_iso}. By employing the trace property, the right hand side is equal to $\frac{1}{\tau}\text{Tr}\left(\mathbf{\Gamma}_{ji}\right)$. Suppose that the e.s.d. of $\mathbf{\Gamma}_{ji}$ converges almost surely to the non-random limiting spectral distribution (l.s.d.) $F_{\mathbf{\Gamma}_{ji}}$ as $K_{j},\tau\to\infty$ with $\frac{K_{j}}{\tau}\to\kappa_{j}$,
$\frac{1}{\tau}\text{Tr}\left(\mathbf{\Gamma}_{ji}\right)=\frac{K_{j}}{\tau}\frac{1}{K_{j}}\text{Tr}\left(\mathbf{\Gamma}_{ji}\right)$
converges almost surely to $\kappa_{j}\mathbb{E}\left[\mathsf{\Gamma}_{ji}\right]$
where $\mathsf{\Gamma}_{ji}$ is a random variable with distribution $F_{\mathbf{\Gamma}_{ji}}$. Consequently, $\upsilon_{ki}\as\bar{\upsilon}_{ki}$, where $\bar{\upsilon}_{ki}$ is defined in \eqref{eq:lim_chest_var}.

\subsection{The large system limit for $A_{ki}$}

From the channel model described in subsection \ref{subsec:Uplink-Training},
$\widehat{\mathbf{h}}_{kii}\sim\mathcal{CN}\left(\mathbf{0},\upsilon_{ki}\mathbf{I}_{N}\right)$. 
By using Lemma \ref{lemma:Lemma1_jse}, $\frac{1}{N}\|\widehat{\mathbf{h}}_{kii}\|^{2}\asymp \upsilon_{ki}$. Since $\upsilon_{ki}\as \bar{\upsilon}_{ki}$, we have
$\frac{1}{N^{2}}A_{ki}\as \rho_{kii}\left(\bar{\upsilon}_{ki}\right)^{2}$.

\subsection{The large system limit for $B_{ki}$}

First let us consider the self interference noise ($\xi_{ki}$). Note that while  $\frac{1}{N}\widehat{\mathbf{h}}_{kii}^{\ct}\widetilde{\mathbf{h}}_{kii}\asymp 0$ by Lemma \ref{lemma:Lemma1_jse}. Consequently, $\frac{1}{N^2}\xi_{ki}\as 0$.
For the noise term, while $\frac{1}{N}\|\widehat{\mathbf{h}}_{kii}\|^{2}\as\bar{\upsilon}_{ki}$, we have  $\frac{1}{N^{2}}\|\widehat{\mathbf{h}}_{kii}\|^{2}\as 0$. Thus, the noise term converges to zero almost surely.

For the intra-cell interference term, it can be written as $\frac{1}{N^{2}}\widehat{\mathbf{h}}_{kii}^{\ct}\mathbf{H}_{ii[k]}\mathbf{\Gamma}_{ii[k]}\mathbf{H}_{ii[k]}^{\ct}\widehat{\mathbf{h}}_{kii}$. Since $\widehat{\mathbf{h}}_{kii}$ is independent of $\mathbf{H}_{ii[k]}$
and $\mathbf{\Gamma}_{ii[k]}$ and by applying Lemma \ref{lemma:Lemma1_jse} and the rank-one perturbation lemma \cite{Couillet_book11}, the intra-cell interference is asymptotically equivalent to $\frac{\bar{\upsilon}_{ki}}{N}\text{Tr}\left(\frac{1}{N}\mathbf{H}_{ii}\mathbf{\Gamma}_{ii}\mathbf{H}_{ii}^{\ct}\right)$. Note that $\frac{1}{N}\text{Tr}\left(\frac{1}{N}\mathbf{H}_{ii}\mathbf{\Gamma}_{ii}\mathbf{H}_{ii}^{\ct}\right)$ is the first moment of the e.s.d. of $\frac{1}{N}\mathbf{H}_{ii}\mathbf{\Gamma}_{ii}\mathbf{H}_{ii}^{\ct}$. It can be shown that (see also \cite[eq. 2.118]{Tulino_fnt04}), this
moment converges almost surely to $\alpha_{i}\mathbb{E}\left[\mathsf{\Gamma}_{ii}\right]$.
Hence, the intra-cell interference power converges to $\alpha_{i}\bar{\upsilon}_{ki}\mathbb{E}\left[\mathsf{\Gamma}_{ii}\right]$
almost surely.

Now, let us consider the inter-cell interference term. The (normalized)
interference from cell $j$ can be written as $\mathcal{I}_{kij}=\frac{1}{N^{2}}\widehat{\mathbf{h}}_{kii}^{\ct}\mathbf{H}_{ji}\mathbf{\Gamma}_{ji}\mathbf{H}_{ji}^{\ct}\widehat{\mathbf{h}}_{kii}$. Let us rewrite \eqref{eq:ch_est_vec}
\begin{align*}
	\widehat{\mathbf{h}}_{kii} & =\frac{\upsilon_{ki}}{\sqrt{\rho_{kii}}}\mathbf{H}_{ji}\mathbf{\Gamma}_{ji}^{\frac{1}{2}}\mathbf{Q}_{j}^{\ct}\mathbf{q}_{ki}+\mathbf{z}_{ki}
\end{align*}
where $\mathbf{z}_{ki}=\upsilon_{ki}\mathbf{h}_{kii}+\frac{\upsilon_{ki}}{\sqrt{\rho_{kii}}}\left(\sum_{l\neq\{i,j\}}^{L}\mathbf{H}_{li}\mathbf{\Gamma}_{li}^{\frac{1}{2}}\mathbf{Q}_{l}^{\ct}\mathbf{q}_{ki}+\hat{\mathbf{n}}_{ki}\right)$.
Let $\mathbf{a}_{c,j}=\frac{\upsilon_{ki}}{\sqrt{\rho_{kii}}}\mathbf{H}_{ji}\mathbf{\Gamma}_{ji}^{\frac{1}{2}}\mathbf{Q}_{j}^{\ct}\mathbf{q}_{ki}$
that represents the (pilot) contamination from cell $j$. Thus,
\begin{align*}
	\mathcal{I}_{kij} & =\frac{1}{N^{2}}\mathbf{z}_{ki}^{\ct}\mathbf{H}_{ji}\mathbf{\Gamma}_{ji}\mathbf{H}_{ji}^{\ct}\mathbf{z}_{ki}+\frac{1}{N^{2}}\mathbf{a}_{c,j}^{\ct}\mathbf{H}_{ji}\mathbf{\Gamma}_{ji}\mathbf{H}_{ji}^{\ct}\mathbf{a}_{c,j}\\
	& \quad+\frac{2}{N^{2}}\Re\left\{ \mathbf{z}_{ki}^{\ct}\mathbf{H}_{ji}\mathbf{\Gamma}_{ji}\mathbf{H}_{ji}^{\ct}\mathbf{a}_{c,j}\right\} .
\end{align*}
It is easy to check that $\mathbf{z}_{ki}$ and $\mathbf{a}_{c,j}$
are (mutually) independent. Furthermore, $\mathbf{z}_{ki}$ and $\mathbf{a}_{c,j}$
are zero mean Gaussian vectors with variances $\psi_{ki}\mathbf{I}_{N}$
with $\psi_{ki}=\upsilon_{ki}^{2}+\frac{\upsilon_{ki}^{2}}{\rho_{kii}}\left(\sigma^{2}+\sum_{l\neq\{i,j\}}^{L}\mathbf{q}_{ki}^{\ct}\mathbf{Q}_{l}\mathbf{\Gamma}_{li}\mathbf{Q}_{l}^{\ct}\mathbf{q}_{ki}\right)$
and $\frac{\upsilon_{ki}^{2}}{\rho_{kii}}\mathbf{q}_{ki}^{\ct}\mathbf{Q}_{j}\mathbf{\Gamma}_{ji}\mathbf{Q}_{j}^{\ct}\mathbf{q}_{ki}$,
respectively. Let us denote $\mathcal{I}_{kij,1}$, $\mathcal{I}_{kij,2}$
and $\mathcal{I}_{kij,3}$ respectively for each term of $\mathcal{I}_{kij}$.
Considering $\mathcal{I}_{kij,1}$, it is obvious that $\mathbf{z}_{ki}$
is independent of $\mathbf{H}_{ji}\mathbf{\Gamma}_{ji}\mathbf{H}_{ji}$. Hence, $\mathcal{I}_{kij,1}\asymp \frac{\psi_{ki}}{N}\TR{\frac{1}{N}\mathbf{H}_{ji}\mathbf{\Gamma}_{ji}\mathbf{H}_{ji}}$.
As shown previously $\mathbf{q}_{ki}^{\ct}\mathbf{Q}_{l}\mathbf{\Gamma}_{li}\mathbf{Q}_{l}^{\ct}\mathbf{q}_{ki}\as\kappa_{l}\mathbb{E}\left[\mathsf{\Gamma}_{li}\right]$. Thus, $\psi_{ki}\as \bar{\psi}_{ki}$ where
\[
\bar{\psi}_{ki}=\frac{\bar{\upsilon}_{ki}^{2}}{\rho_{kii}}\left(\rho_{kii}+\sigma^{2}+\sum_{l\neq\{i,j\}}^{L}\kappa_{l}\mathbb{E}\left[\mathsf{\Gamma}_{li}\right]\right).
\]
Moreover, $\frac{1}{N}\TR{\frac{1}{N}\mathbf{H}_{ji}\mathbf{\Gamma}_{ji}\mathbf{H}_{ji}}$  converges to $\alpha_{j}\mathbb{E}\left[\mathsf{\Gamma}_{ji}\right]$ almost surely. Thus,
\[
\mathcal{I}_{kij,1}\as \frac{\alpha_{j}\bar{\upsilon}_{ki}^{2} \mathbb{E}\left[\mathsf{\Gamma}_{ji}\right]}{\rho_{kii}}\left(\rho_{kii}+\sigma^{2}+\sum_{l\neq\{i,j\}}^{L}\kappa_{l}\mathbb{E}\left[\mathsf{\Gamma}_{li}\right]\right).
\]

For $\mathcal{I}_{kij,2}$, we have 
\begin{align*}
\mathcal{I}_{kij,2}&=\frac{\upsilon_{ki}^{2}}{\rho_{kii}}\left(\frac{1}{N^{2}}\mathbf{q}_{ki}^{\ct}\mathbf{Q}_{j}\mathbf{\Gamma}_{ji}^{\frac{1}{2}}\mathbf{H}_{ji}^{\ct}\mathbf{H}_{ji}\mathbf{\Gamma}_{ji}\mathbf{H}_{ji}^{\ct}\mathbf{H}_{ji}\mathbf{\Gamma}_{ji}^{\frac{1}{2}}\mathbf{Q}_{j}^{\ct}\mathbf{q}_{ki}\right)\\
&\asymp \frac{\upsilon_{ki}^{2}}{\rho_{kii}}\times\frac{N}{\tau}\times\frac{1}{N} \text{Tr}\left(\left(\frac{1}{N}\mathbf{H}_{ji}\mathbf{\Gamma}_{ji}\mathbf{H}_{ji}^{\ct}\right)^{2}\right).
\end{align*}
The last term in the last line is the second moment of the e.s.d. of $\frac{1}{N}\mathbf{H}_{ji}\mathbf{\Gamma}_{ji}\mathbf{H}_{ji}^{\ct}$.
It can be shown that (see also \cite[eq. 2.118]{Tulino_fnt04})
it  converges almost surely to $\alpha_{j}\mathbb{E}\left[\mathsf{\Gamma}_{ji}^{2}\right]+\alpha_{j}^{2}\left(\mathbb{E}\left[\mathsf{\Gamma}_{ji}\right]\right)^{2}$.
By writing $\frac{N}{\tau}=\frac{K_{j}}{\tau}\frac{N}{K_{j}}\to\frac{\kappa_j}{\alpha_j}$, we obtain
\[
\mathcal{I}_{kij,2}\as\frac{\bar{\upsilon}_{ki}^{2}}{\rho_{kii}}\kappa_{j}\left(\mathbb{E}\left[\mathsf{\Gamma}_{ji}^{2}\right]+\alpha_{j}\left(\mathbb{E}\left[\mathsf{\Gamma}_{ji}\right]\right)^{2}\right).
\]

Let us consider the term $\frac{1}{N^{2}} \mathbf{z}_{ki}^{\ct}\mathbf{H}_{ji}\mathbf{\Gamma}_{ji}\mathbf{H}_{ji}^{\ct}\mathbf{a}_{c,j}$ in 
 $\mathcal{I}_{kij,3}$. It can be written as 
\begin{align*}
	 &\frac{\upsilon_{ki}}{N^2\sqrt{\rho_{kii}}}\mathbf{z}_{ki}^\ct\mathbf{H}_{ji}\mathbf{\Gamma}_{ji}\mathbf{H}_{ji}^{\ct}\mathbf{H}_{ji}\mathbf{\Gamma}_{ji}^{\frac{1}{2}}\mathbf{Q}_{j}^{\ct}\mathbf{q}_{ki} \\
	 &\quad= \frac{\upsilon_{ki}}{N^2\sqrt{\rho_{kii}}}\sum_{m=1}^{K_j}\sqrt{\rho_{mji}}(\mathbf{q}_{mj}^\ct\mathbf{q}_{ki})\mathbf{z}_{ki}^\ct\mathbf{H}_{ji}\mathbf{\Gamma}_{ji}\mathbf{H}_{ji}^{\ct}\mathbf{h}_{mji}\\
	 &\quad=\frac{\upsilon_{ki}}{N\sqrt{\rho_{kii}}}\sum_{m=1}^{K_j}\sqrt{\rho_{mji}}(\mathbf{q}_{mj}^\ct\mathbf{q}_{ki})\frac{1}{N}\mathbf{z}_{ki}^\ct\mathbf{H}_{ji[m]}\mathbf{\Gamma}_{ji[m]}\mathbf{H}_{ji[m]}^{\ct}\mathbf{h}_{mji}\\
	 &\qquad + \frac{\upsilon_{ki}}{N\sqrt{\rho_{kii}}}\sum_{m=1}^{K_j}\rho_{mji}^{\frac{3}{2}}(\mathbf{q}_{mj}^\ct\mathbf{q}_{ki})\frac{\mathbf{z}_{ki}^\ct\mathbf{h}_{mji}\mathbf{h}_{mji}^{\ct}\mathbf{h}_{mji}}{N}. 
\end{align*}
It can be checked that all the summands of the last equation are asymptotically equivalent to zero. Therefore, by using Lemma~\ref{lemma:av_seq_lim}, $\mathcal{I}_{kij,3}\as 0$.
 
Summing up over all interfering cells, the limiting result for the total inter-cell
interference $\sum_{j\neq i}\mathcal{I}_{kij}$ is 
\begin{align*}
	& \frac{\bar{\upsilon}_{ki}^{2}}{\rho_{kii}}\sum_{j\neq i}^{L}\kappa_{j}\mathbb{E}\left[\mathsf{\Gamma}_{ji}^{2}\right]\\
	& \qquad+\frac{\bar{\upsilon}_{ki}^{2}}{\rho_{kii}}\left(\rho_{kii}+\sum_{l\neq i}^{L}\kappa_{l}\mathbb{E}\left[\mathsf{\Gamma}_{li}\right]+\sigma^{2}\right)\sum_{j\neq i}^{L}\alpha_{j}\mathbb{E}\left[\mathsf{\Gamma}_{ji}\right].
\end{align*}
Using the definition for $\bar{\upsilon}_{ki}$ in \eqref{eq:lim_chest_var},
the second term can be simplified to $\bar{\upsilon}_{ki}\sum_{j\neq i}^{L}\alpha_{j}\mathbb{E}\left[\mathsf{\Gamma}_{ji}\right]$.
Thus,
\[
\sum_{j\neq i}\mathcal{I}_{kij}\as\bar{\upsilon}_{ki}\left(\sum_{j\neq i}^{L}\alpha_{j}\mathbb{E}\left[\mathsf{\Gamma}_{ji}\right]+\frac{\bar{\upsilon}_{ki}}{\rho_{kii}}\sum_{j\neq i}^{L}\kappa_{j}\mathbb{E}\left[\mathsf{\Gamma}_{ji}^{2}\right]\right).
\]
Bringing up the limiting results for each term of $B_k$, we obtain
\[
\frac{1}{N^2}B_{ki}\as\bar{\upsilon}_{ki}\left(\sum_{j=1}^{L}\alpha_{j}\mathbb{E}\left[\mathsf{\Gamma}_{ji}\right]+\frac{\bar{\upsilon}_{ki}}{\rho_{kii}}\sum_{j\neq i}^{L}\kappa_{j}\mathbb{E}\left[\mathsf{\Gamma}_{ji}^{2}\right]\right).
\]
By putting the large system results for $A_{ki}$ and $B_{ki}$, \eqref{eq:asinr_ul} follows immediately and this completes the proof.
\bibliographystyle{IEEEtran}
\bibliography{massive}
 
\end{document}